\documentclass[3p]{elsarticle}

\usepackage{hyperref}
\usepackage[utf8]{inputenc}
\usepackage[T1]{fontenc}
\usepackage{url}
\usepackage{amsmath}
\usepackage{amssymb}
\usepackage{mathtools}
\usepackage{graphicx}
\usepackage{paralist}
\usepackage{textcomp}
\usepackage{gensymb}
\usepackage{nicefrac}
\usepackage[dvipsnames]{xcolor}

\journal{NeuroImage}

\begin{document}

\begin{frontmatter}

%
\pagestyle{headings}  
%
\title{Multi-scale Graph-based Grading for Alzheimer's Disease Prediction}

\author[UB]{Kilian Hett}
\author[UB]{Vinh-Thong Ta}
\author[UPB]{Jos\'e V. Manj\'on}
\author[UB]{Pierrick Coup\'e}
\author{Alzheimer's Disease Neuroimaging Initiative\fnref{adniFootnote}}

\fntext[adniFootnote]{Data used in preparation of this article were obtained from the Alzheimer's Disease Neuroimaging Initiative (ADNI) database (\url{adni.loni.usc.edu}).  As such, the investigators within the ADNI contributed to the design and implementation of ADNI and/or provided data but did not participate in analysis or writing of this report. A complete listing of ADNI investigators can be found at: \url{http://adni.loni.usc.edu/wp-content/uploads/how_to_apply/ADNI_Acknowledgement_List.pdf}.}

\address[UB]{CNRS, Univ. Bordeaux, Bordeaux INP, LABRI, UMR5800, PICTURA, F-33400 Talence, France}
\address[UPB]{Universitat Polit\`ecnia de Val\`encia, ITACA, 46022 Valencia, Spain}

\biboptions{longnamesfirst,semicolon}

\begin{abstract}
The prediction of subjects with mild cognitive impairment (MCI) who will progress to Alzheimer's disease (AD) is clinically relevant, and may above all have a significant impact on accelerate the development of new treatments. In this paper, we present a new MRI-based biomarker that enables us to predict conversion of MCI subjects to AD accurately. In order to better capture the AD signature, we introduce two main contributions. First, we present a new graph-based grading framework to combine inter-subject similarity features and intra-subject variability features. This framework involves patch-based grading of anatomical structures and graph-based modeling of structure alteration relationships. Second, we propose an innovative multiscale brain analysis to capture alterations caused by AD at different anatomical levels. Based on a cascade of classifiers, this multiscale approach enables the analysis of alterations of whole brain structures and hippocampus subfields at the same time. During our experiments using the ADNI-1 dataset, the proposed multiscale graph-based grading method obtained an area under the curve (AUC) of 81\% to predict conversion of MCI subjects to AD within three years. Moreover, when combined with cognitive scores, the proposed method obtained 85\% of AUC. These results are competitive in comparison to state-of-the-art methods evaluated on the same dataset.
\end{abstract}


\end{frontmatter}

\section{Introduction}

Alzheimer's disease (AD) is the most prevalent dementia affecting elderly people \citep{petrella2003neuroimaging}. According to the World Health Organization, the number of patients with AD will double in 20 years. AD is a serious condition characterized by an irreversible neurodegenerative process that causes mental dysfunctions such as longterm memory loss, language impairment, disorientation, change in personality, and finally death \citep{alzheimer20152015}. This disease is characterized by an accumulation of beta-amyloid plaques and neurofibrillary tangles composed of tau amyloid fibrils \citep{hardy2006alzheimer} leading to synapse and neuronal losses. To date, no known therapy has been able to stop or slow down the progression of AD. Moreover, neuroimaging studies have revealed that brain changes occur decades before the diagnosis is established \cite{coupe2015detection,coupe2019lifespan}. Thus, when the diagnosis is made, the pathological load is already high \citep{decarli2003mild}. 

Indeed, before the diagnosis is established the patient is already suffering from amnesic mild cognitive impairment (MCI). MCI is considered a prodromal phase of AD. The clinical symptoms of MCI are slight but a decrease of cognitive abilities is measurable. Previous studies have suggested that approximately 12\% of subjects suffering from MCI progress to AD in the four years following the first symptoms \citep{petersen1999mild}. Therefore, although MCI subjects present a high risk of AD development, subjects suffering from MCI can remain stable (\emph{i.e.}, do not convert to AD) or even recover normal cognitive status. The early prediction of the subjects suffering from MCI symptoms who will convert to AD is thus crucial. This can improve the effectiveness of the future therapies by reducing the brain changes before the therapy starts. Also, the prediction of conversion to AD can accelerate the development of new therapies by making the subject selection more accurate. This would decrease the cost of clinical trials and enable more accurate clinical studies.

With the improvement of medical imaging techniques such as magnetic resonance imaging (MRI), many methods have been developed to increase the ability of computer-aided diagnosis systems to help early AD detection \citep{arbabshirani2017single,rathore2017review}. These methods can be grouped into two categories related to how they analyze AD alterations. 
	
On the one hand, methods have been developed to study the inter-subject similarity between individuals from different groups that represent specific disease severities. Among these approaches,  a popular method to estimate similarity at a voxel scale is the voxel-based morphometry (VBM) \citep{ashburner2000voxel,bron2015standardized,moradi2015machine}. Methods based on region of interest (ROI) have also been proposed. A widely used approach is based on a volumetric measurement of gray matter within brain structures \citep{bron2015standardized,ledig2018structural}. Other ROI-based methods such as thickness measurement have been developed to capture the variations of gray matter along the cerebral cortex \citep{wolz2011multi, wee2013prediction}. Among advanced methods, patch-based grading (PBG) framework has been proposed to capture subtler alterations caused by the pathology. Indeed, PBG method has demonstrated state-of-the-art performance to detect alterations of hippocampus \citep{coupe2012scoring,hett2018adaptive}. This framework has also been extended to perform a whole brain analysis \cite{tong2017novel}. This extension has shown competitive performance for AD prediction especially compared to other approaches based on deep-learning architectures \cite{basaia2018automated,lian2018hierarchical}. 

On the other hand, several methods have been proposed to capture the intra-subject variability. Such methods assume that AD does not occur at isolated areas but in several inter-related regions. Although similarity-based biomarkers provide helpful tools for detecting the first signs of AD, the structural alterations leading to cognitive decline are not homogeneous within a given subject. Therefore, intra-subject variability features could encode relevant information. Some methods proposed to capture the relationship of spread cortical atrophy with a network-based framework \cite{wee2013prediction}. Other approaches estimate inter-regional correlation of brain tissue volumes \cite{zhou2011hierarchical}. A study has proposed a generic framework that embeds spatial and anatomical priors within a graph model. This method extracts inter-subject variability from different features (for instance, voxel-based and cortical thickness) and various MRI modalities \cite{cuingnet2013spatial}. Recently, convolutional neural networks (CNN) have been used to capture relationships between anatomical structures volumes \cite{suk2017deep}, and cortical thickness \cite{wee2019cortical}. It is interesting to note that methods based on inter-subject similarities and intra-subject variability have performed similarly for AD prediction. 

All these elements indicate that inter-subject similarity and intra-subject variability features provide important information for predicting the subject's conversion. Consequently, our first contribution is the development of a new method that efficiently combines inter-subject similarities estimated with a patch-based grading approach and intra-subjects' variability modeled by a graph-based approach. We applied our new method to two different anatomical scales: hippocampal subfields and whole brain structures. The experiments carried out show an increase in prediction performances for both anatomical scales. This demonstrates the generic nature of our new method. The second contribution presented in this paper is the development of a novel method based on a cascade of classifiers to efficiently and simultaneously combine information related to hippocampal subfields and whole brain structures alterations. Our multi-scale graph-based grading method demonstrates competitive performance with an area under the curve (AUC) of 81\% for AD prediction. Moreover, when combined with cognitive scores, the proposed method obtained 85\% of AUC.


\section{Materials}

\subsection{Dataset}
Data used in this work were obtained from the Alzheimer's Disease Neuroimaging Initiative (ADNI) dataset\footnote[1]{\url{http://adni.loni.ucla.edu}}. ADNI is a North American campaign launched in 2003 with aims to provide MRI, positron emission tomography scans, clinical neurological measures and other biomarkers. We use baseline T1-weighted (T1w) MRI of the ADNI1 phase that has been proposed in \citep{tong2017novel}. This dataset includes AD patients, subjects with mild cognitive impairment (MCI) and cognitive normal (CN) subjects. MCI is a presymptomatic phase of AD composed of subjects who have abnormal memory dysfunctions. In our experiments we consider two groups of MCI. The first group is composed of patients who have stable MCI (sMCI) and the second one is composed of patients having MCI symptoms at the baseline and then converted to AD in the following 36 months. This group is named progressive MCI (pMCI). The information of the dataset used in our work is summarized in Table \ref{tab:dataset}. Moreover, the dataset and the code developed during this study are available from the corresponding author on reasonable request.

\begin{table*}[!h]
\begin{center}
\caption[Description of the dataset used in this work]{Description of the dataset used in this work. Data are provided by ADNI.}\label{tab:dataset}
\begin{tabular}{@{\hspace{0.6cm}} l @{\hspace{0.6cm}} c @{\hspace{0.6cm}} c @{\hspace{0.6cm}} c @{\hspace{0.6cm}} c @{\hspace{0.6cm}} c @{\hspace{0.6cm}}} \hline
 & CN & sMCI & pMCI & AD & P value  \\ \hline
Number  	& 213 & 90 & 126 & 130 & \\
Ages (years) 	& $75.7 \pm 5.0$ 	& $74.9 \pm 7.5$ 	& $73.7 \pm 7.0$ 	& $74.1 \pm 7.7$ 	& {$p$ = 0.63$^a$} \\
Sex (M/F) 	& $108/105$ 		& $58/32$ 		& $68/58$ 		&  $64/66$ 		& {$\chi^2$=5.29, $p$ = 0.15$^b$} \\ 
MMSE 		& $29.1 \pm 1.0$ 	& $27.6 \pm 1.7$ 	& $26.5 \pm 1.6$ 	& $23.5 \pm 1.9$ 	& {$p$  < 0.01$^{a*}$}\\ 
CDR-SB		& 3.5 $\pm$ 2.7		& 4.5 $\pm$ 2.3 	& 4.8 $\pm$ 2.1  	& 4.7  $\pm$ 1.9 	&  {$p$  < 0.01$^{a*}$} \\ 
RAVLT		& 45.4 $\pm$ 9.7	& 35.5 $\pm$ 10.2 	& 27.7 $\pm$ 8.9  	& 24.6 $\pm$ 7.0 	&  {$p$  < 0.01$^{a*}$} \\ 
FAQ		& 8.4  $\pm$ 4.4	& 13.3 $\pm$ 5.4 	& 20.2 $\pm$ 6.7  	& 30.0 $\pm$ 9.0 	&  {$p$  < 0.01$^{a*}$} \\ 
ADAS11		& 5.2  $\pm$ 3.0	& 8.1  $\pm$ 3.6 	& 12.5 $\pm$ 4.9  	& 20.2 $\pm$ 7.6 	&  {$p$  < 0.01$^{a*}$} \\ 
ADAS13		& 0.2 $\pm$ 0.9		& 2.3 $\pm$ 3.7 	& 4.3 $\pm$ 4.8   	& 14.6 $\pm$ 6.6 	&  {$p$  < 0.01$^{a*}$} \\  \hline
\end{tabular}
\end{center}
 \hspace{0.5cm} {$^*$ Significant at p < 0.05.}
 
 \hspace{0.5cm} {$^a$ Chi-square test (df = 3).}
 
 \hspace{0.5cm} {$^b$ Kruskal–Wallis test (df = 3).}
\end{table*}


\subsection{Preprocessing}
The data are preprocessed using the following steps: (1) denoising using a spatially adaptive non-local means filter \citep{manjon2010adaptive}, (2) inhomogeneity correction using N4 method \citep{tustison2010n4itk}, (3) low-dimensional non-linear registration to MNI152 space using ANTS software \citep{avants2011reproducible}, (4) intensity standardization. All the following experiments have been carried out with images in the MNI space.

\section{Method}

\subsection{Method overview}
\begin{figure*}[!ht]
\centering
\includegraphics[width=1\linewidth]{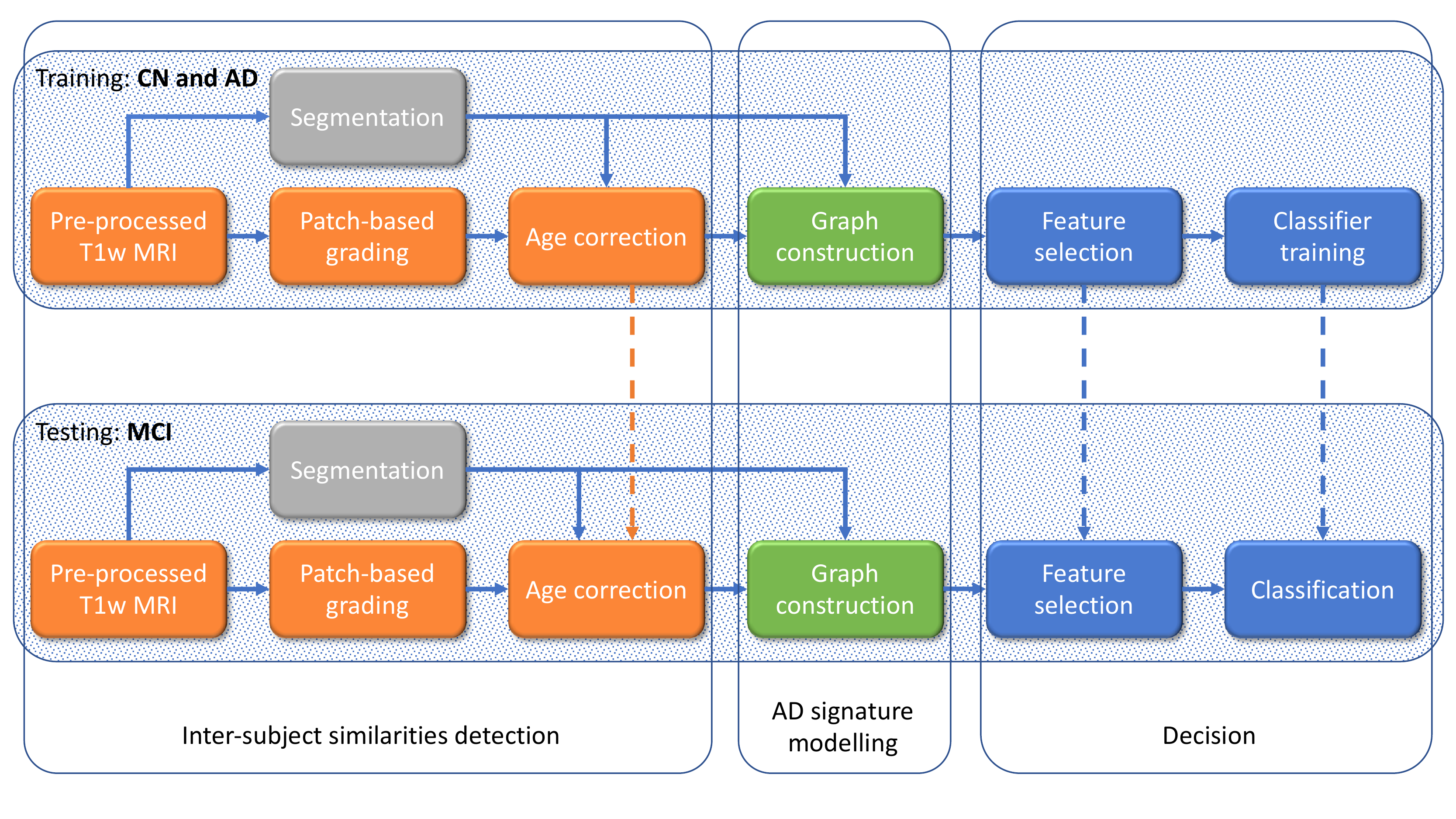}
\caption{Pipeline of the proposed graph-based grading method. PBG is computed using CN and AD training groups. CN group is also used to correct the bias related to age. Then, this estimation is applied to AD and MCI subjects. Afterwards, the graph is constructed, and the feature selection is trained on CN and AD and then is applied to CN, AD and MCI. Finally the classifier is trained with CN and AD.}
\label{fig:gsg_workflow}
\end{figure*}

As illustrated in Figure~\ref{fig:gsg_workflow}, our graph of structure grading method, that combines inter-subjects' similarities and intra-subjects' variability, is composed of several steps. First, a segmentation of the structures of interest is conducted on the input images. Then, a patch-based grading (PBG) approach is carried out over every segmented structures (\emph{e.g.}, hippocampal subfields and brain structures). Two different alterations impacting the brain structures are captured with PBG methods: the changes caused by normal aging \citep{koikkalainen2012improved} and the alterations caused by the progression of AD. Therefore, at each voxel, the grading values are age-corrected to avoid bias due to normal aging. After the patch-based grading maps are age-corrected, we construct an undirected graph to model the topology of alterations caused by Alzheimer's disease. This results in a high dimensional feature vector. Consequently, to reduce the dimensionality of the feature vector computed by our graph-based method, we use an elastic net that provides a sparse representation of the most discriminative elements of our graph (\emph{i.e.}, edges and vertices). We use only the most discriminative features of our graph as the input to a random forest method which predicts the subject's conversion.

\subsection{Segmentation}

First, to enable analysis of the alterations that occur over different brain structures, segmentation using a non-local label fusion \citep{giraud2016optimized} and a systematic error correction \citep{wang2011learning} at two different anatomical scales are performed, the hippocampal subfields and the whole brain structures. 

Segmentation of hippocampal subfields was performed with HIPS, which is a method based on a combination of non-linear registration and patch-based label fusion \citep{romero2017hips}. This method uses a training library based on a dataset composed of high resolution T1w images manually labeled according to the protocol proposed in \citep{winterburn2013novel}. To perform the segmentation, the ADNI images were up-sampled with a local adaptive super-resolution method to fit in the training image resolution \citep{coupe2013collaborative}. The method provides automatic segmentation of hippocampal subfields grouped into 5 labels: Subiculum, CA1SP, CA1SR-L-M, CA2-3 and CA4/DG. Afterwards, the segmentation maps obtained on the up-sampled T1w images were down-sampled to fit in the previous MNI space resolution.

Whole brain structures have been labeled with a patch-based multi-template segmentation \cite{manjon2016volbrain}. This method has been performed using 35 images manually labeled  by Neuromorphometrics, Inc. \footnote{\url{http://Neuromorphometrics.com}} following the brain-COLOR labeling protocol composed of 133 structures.

 Finally, visual quality control was conducted to remove all incorrect segmentations from the dataset. Moreover, to prevent any bias in the dataset, the pathological status of each subject was hidden during the entire quality control process. 

\subsection{Patch-based grading}
Once the images are segmented, a patch-based grading of the entire brain is performed using the method described in \cite{hett2018adaptive}. Such a method was first proposed to detect hippocampus structural alterations with a new scale of analysis \cite{coupe2012simultaneous,coupe2012scoring}. The patch-based grading approach provides the probability that the disease has impacted the underlying structure at each voxel. This probability is estimated via an inter-subject similarity measurement derived from a non-local approach. 

The method begins with building a training library $T$ from two datasets of images: one with images from CN subjects and the other one from AD patients. Then, for each voxel $x_i$ of the region of interest in the considered subject $x$, the PBG method produces a weak classifier denoted $g_{x_i}$. This weak classifier provides a surrogate for the pathological grading at the considered position $i$. The weak classifier is computed using a measurement of the similarity between the patch $P_{x_i}$ surrounding the voxel $x_i$ belonging to the image under study and a set $K_{x_i} = \{ P_{t_{j}} \}$ of the closest patches $P_{t_j}$, surrounding the voxel $t_j$, extracted from the template $t \in T$. The grading value $g_{x_i}$ at $x_i$ is defined as:

\begin{equation}
g_{x_i} = \frac{\sum_{t_{j} \in K_{x_i}} w(P_{x_i},P_{t_{j}}) p_t}{\sum_{t_{j} \in K_{x_i}} w(P_{x_i},P_{t_{j}})} \label{eq:grade}
\end{equation} 

where $w(x_{i},t_{j})$ is the weight assigned to the pathological status $p_t$ of the training image $t$. We estimate $w$ as:

\begin{equation}
w(P_{x_{i}},P_{t_{j}}) = \exp \Big( - \tfrac{||P_{x_{i}} - P_{t_{j}}||^2_2}{h^2}  \Big) \label{eq:weight}
\end{equation}

where $h=\min ||P_{x_i} - P_{t_{j}}||^2_2 + \epsilon$ and $\epsilon \to 0$. The pathological status $p_t$ is set to $-1$ for patches extracted from AD patients and to $1$ for patches extracted from CN subjects. Therefore, the PBG method provides a score representing an estimate of the alterations caused by AD at each voxel. Consequently, cerebral tissues strongly altered by AD have scores close to $-1$ while healthy tissues have scores close to $1$.

\subsection{Graph construction}
Once structure alterations are estimated using patch-based grading, we can model intra-subject variability for each subject using a graph to better capture the AD signature. Indeed, within the last decade, graph modeling has been widely used for its ability to capture the patterns of different diseases. This is achieved by encoding the relationships of abnormalities between different structures in the edges of the graph. Most of the proposed methods estimate the degree of correlation between two different structures for each edge of the graph. Furthermore, graph modeling can also model inter-subject similarity, by independently encoding the abnormality of each structure in the vertices measurement. Consequently, we proposed a graph-based grading approach that uses a graph model to combine inter-subject similarities computed with the PBG and intra-subjects' variability which is computed with a distribution of the grading value distributions within each structure.

In our graph-based grading method, the segmentation maps are used to fuse grading values into each ROI, and to build our graph. We define an undirected graph $G=(V, E, \gamma, \omega)$, where $V=\{v_1,...,v_N\}$ is the set of vertices for the $N$ considered brain structures and $E=V \times V$ is the set of edges. In our work, the vertices are the mean of the grading values for a given structure while the edges are based on grading distribution distances between two structures. 

To this end, the probability distributions of PBG values are estimated with a histogram $H_v$ for each structure $v$. The number of bins is computed with Sturge's rule \citep{sturges1926choice}. For each vertex we assign a function $\gamma:V\rightarrow \mathbb{R}$ defined as $\gamma(v) = \mu_{H_v}$, where $\mu_{H_v}$ is the mean of $H_v$. For each edge we assign a weight given by the function $\omega:E\rightarrow \mathbb{R}$ defined as follows:

\begin{equation}
	\omega(v_i,v_j) = \exp ( -W(H_{v_i},H_{v_j})^2 / \sigma^2  )
\end{equation}

where $W$ is the Wasserstein distance with $L_1$ norm \citep{rubner2000earth} that showed best performance during our experiments. Wasserstein distance between two histograms is defined as the minimization of the following equation,

\begin{equation}
	 W(H_{v_i},H_{v_j},F) = \min_{F=\{f_{k;l}\}} \sum_{k,l} f_{k;l} d_{k;l}
\end{equation}

subject to, 

\begin{equation}
\begin{aligned}[c]
    \sum_{k \in I} f_{k;l} = p_{k} & \text{ \hspace{0.5cm}} \forall k \in I \\
    \sum_{l \in I} f_{k;l} = q_{l} & \text{ \hspace{0.5cm}} \forall l \in I \\
    f_{k;l} \geq 0 &  \text{ \hspace{0.5cm}} \forall (k;l) \in J \\
\end{aligned}
\end{equation}

where $I = \{k | 1\leq k \leq m \}$ is the index set for bins, $H_{v_i}=\{p_k | k \in I\}$ and $H_{v_j}=\{q_k | k \in I\}$ are the two normalized histograms. $J=\{(k,l) | k \in I, l \in I\}$ is the set for flows, and $d_{k;l} = ||k - l||_p$ is the group distance defined by a $L_p$ norm. As described above, in our experiment we used the $L_1$-norm.

\subsection{Selection of discriminant graph components}
Completion of the previous step results in a high-dimensional feature vector. Moreover, all the components computed from the graph-based grading method do not have the same significance. For instance, some structures and some alterations relationship are not discriminant to detect Alzheimer's disease. 

Consequently, in this work we used the elastic net regression method that provides a sparse representation of the most discriminating edges and vertices, and thus reducing the feature dimensionality by capturing the key structures and the key relationships between the different brain structures (see Figure~\ref{fig:gsg_workflow}). Indeed, it has been demonstrated that combining the $L1$ and $L2$ norms takes into account possible inter-feature correlation while imposing sparsity \citep{zou2005regularization}. Finally, after normalization, the resulting feature vector is given as the input of the feature selection, defined as the minimization of the following equation:

\begin{equation}
 \hat{\beta} = \underset{\beta}{\min} \frac{1}{2} || X\beta - y ||^2_2 + \rho ||\beta||^2_2 + \lambda ||\beta||_1
\end{equation}

where $\hat{\beta}$ is a sparse vector that represents the regression coefficients and $X$ is a matrix with rows corresponding to the subjects and columns corresponding to the features, including: the vertices, the edges or a concatenation of both for the full graph of grading feature vector. $\rho$ and $\lambda$ are the regularization hyper-parameters set to balance the sparsity and the correlation inter-feature. Finally, $y$ represents the pathological status of each patient.

\subsection{Application to different anatomical scales}
In our experiments two different anatomical scales have been considered. First, as presented in \citep{hett2018ghsg}, we applied our graph of structure grading method within a definition of the hippocampal subfields. A histogram is computed to estimate the probability distribution of the grading values for each hippocampal subfields. Thus, $GG_{subfields}=(V, E, \Gamma, \Omega)$, represents the graph of the hippocampal subfields grading. The vertices $V$ represent alteration of hippocampal subfields measured with patch-based grading, and the edges $E$ represent the relationship between hippocampal subfield alterations embedded in graph modeling.

Second, we applied our graph-based approach to a whole brain parcelization. Here, the histograms are computed to estimate the probability distribution of the grading values within each brain structure as proposed in \citep{hett2018gbsg}. Thus, for this second anatomical scale of analysis, $GG_{brain}=(V, E, \Gamma, \Omega)$ represents the graph of brain structure grading. The vertices $V$ represent measures of alteration of brain structures and the edges $E$ represent the alteration relationship between two brain structures.

\subsection{Multi-scale graph-based grading}
To combine multiple anatomical scales (for instance, brain structures and hippocampal subfields), we developed a multi-scale graph-based grading ($MGG$) approach. 

\begin{figure*}[!ht]
\centering
\includegraphics[width=1\linewidth]{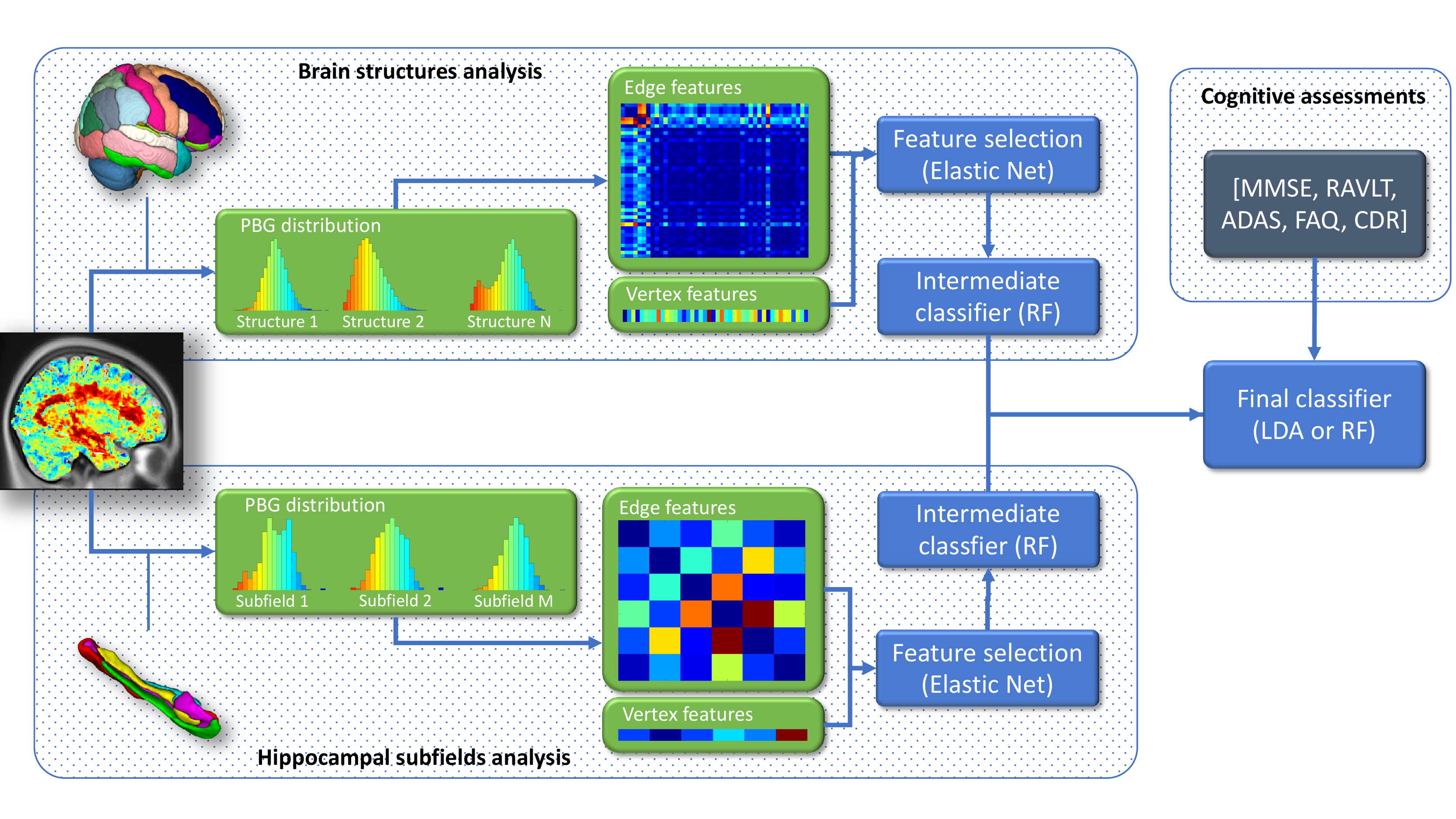}
\caption{Schema of the proposed multi-scale graph-based grading method. First, the segmentation maps are used to aggregate grading values. Our method computes a histogram for each structure/subfield. Once the graphs are built, an elastic net is computed to select the most discriminating graph features for each anatomical scale. A first layer of random forest classifiers are computed to estimate \emph{a posteriori} probabilities. Finally, a linear classifier is trained with the \emph{a posteriori} probabilities from each anatomical scale to compute the final decision. A random forest classifier replaces the linear classifier for the multimodal experiments to deal with the feature heterogeneity resulting from the concatenation of \emph{a posteriori} probabilities and cognitive scores.}
\label{fig:multiple_graph}
\end{figure*}

This method is based on a cascade of classifiers. In this approach, a graph of brain structures and a graph of hippocampal subfields are computed separately as has been described in the previous sections (see Figure~\ref{fig:multiple_graph}). The elastic net regression method is used to select the most discriminating features of each graph. Afterward, a first layer of RF classifier is used to compute both \emph{a posteriori} probabilities $P(p_t|X_{GG_{brain}})$ and $P(p_t|X_{GG_{subfields}})$ for whole brain and hippocampal subfields, respectively. Here, $p_t$ represents the pathological status of the subject under study, while $X_{GG_{brain}}$ and $X_{GG_{subfields}}$ represent the selected features of $GG_{brain}$ and $GG_{subfields}$ models respectively. Finally, these \emph{a posteriori} probabilities are used as the input of a linear classifier to make the final decision.
 
 In addition of this new method, we also proposed a straightforward extension of our graph-based grading method. This approach results in the concatenation of $GG_{brain}$ and $GG_{subfields}$ features into a single feature vector before the feature selection step.

\subsection{Combination with cognitive tests}
It has been shown in previous works that MRI-based biomarkers can be complementary to cognitive assessments used in clinical routines \cite{tong2017novel,samper2019reproducible}.

Therefore, in addition of studying the efficiency of our novel imaging-based biomarkers, a study of the complementarity of our proposed method with cognitive scores has also been conducted. In this work, we have considered different cognitive scores such as MMSE, CDR-SB, RAVLT, FAQ, ADAS11, and ADAS13 cognitive tests. The cognitive scores are concatenated into a vector of cognitive features and normalized by a z-score. Finally, a concatenation of normalized cognitive scores and graph-based features are used as inputs of the final classifier as illustrated in Figure~\ref{fig:multiple_graph}.

\subsection{Details of implementation}
The most similar patches were extracted with a patch-match method \citep{giraud2016optimized}. We used the grading method proposed in \citep{hett2018adaptive}, with the same parameters for the size of the patches and $K_{x_i}$. The effect of age has been corrected using a linear regression estimated on CN population \citep{koikkalainen2012improved}.

The elastic net feature selection has been computed with the SLEP package \citep{liu2009slep}. The two parameters $\lambda$ and $\rho$ have been set up with a grid search method resulting in $\lambda = 0.08$ and $\rho = 0.04$ being the best parameters for the experiment. The classifications were obtained using a random forest (RF) as implemented in~\footnote{\url{http://code.google.com/p/randomforest-matlab}}. In our experiments, we used the Gini index as impurity criterion. RF has also two parameters, the numbers of three $N$ and the number of randomly selected features $T$. These two parameters was set as follows, $N=500$ and $T=4$. A linear discriminant analysis (LDA) classifier has been used to compute the final decision for the multi-scale graph-based grading approach. However, a random forest classifier replaces the linear classifier for the multimodal experiments to deal with the feature heterogeneity resulting from the concatenation of \emph{a posteriori} probabilities and cognitive scores. All features were normalized using z-scores before the selection and classification steps. 

In our experiments, we performed sMCI versus pMCI and CN versus AD classifications. For sMCI versus pMCI classification, the elastic net feature selection and the classifiers were trained with CN and AD. Indeed, as shown in \citep{tong2017novel}, the use of CN and AD to train the feature selection method and the classifier enables to better discrimination between sMCI and pMCI subjects. Furthermore, this technique also limits bias and the overfitting problem and does not require a cross validation step. Finally, to estimate the inner-variability of the RF, 100 runs were performed. A stratified 10-folds cross-validation procedure has been conducted for the comparison of CN versus AD. Mean area under curve (AUC), accuracy (ACC), balanced accuracy (BACC), sensitivity (SEN), and specificity (SPE) are provided for each experiment.

\section{Results}
In this section, to evaluate the performance of the graph-based grading method, we first propose a comparison of the prediction and detection accuracy of the different graph components. Afterwards, we apply our method within the hippocampal subfields and the whole brain structures (see Table~\ref{tab:results1}). Moreover, we evaluate the proposed approach to combine different anatomical scales (see Table~\ref{tab:results_multigraph}). Then, we evaluate the complementarity of our image-based biomarker and the cognitive scores that are usually used in clinical routines (see Table~\ref{tab:cognitivescore_multigraph}). Finally, we compare the performance of our method with state-of-the-art methods for early detection of Alzheimer's disease (see Table~\ref{tab:comparisons1_multigraph} and \ref{tab:MultiSourceClass}).

%

\subsection{Graph of hippocampal subfields}

\begin{figure*}[!ht]
\centering
\includegraphics[width=1\linewidth]{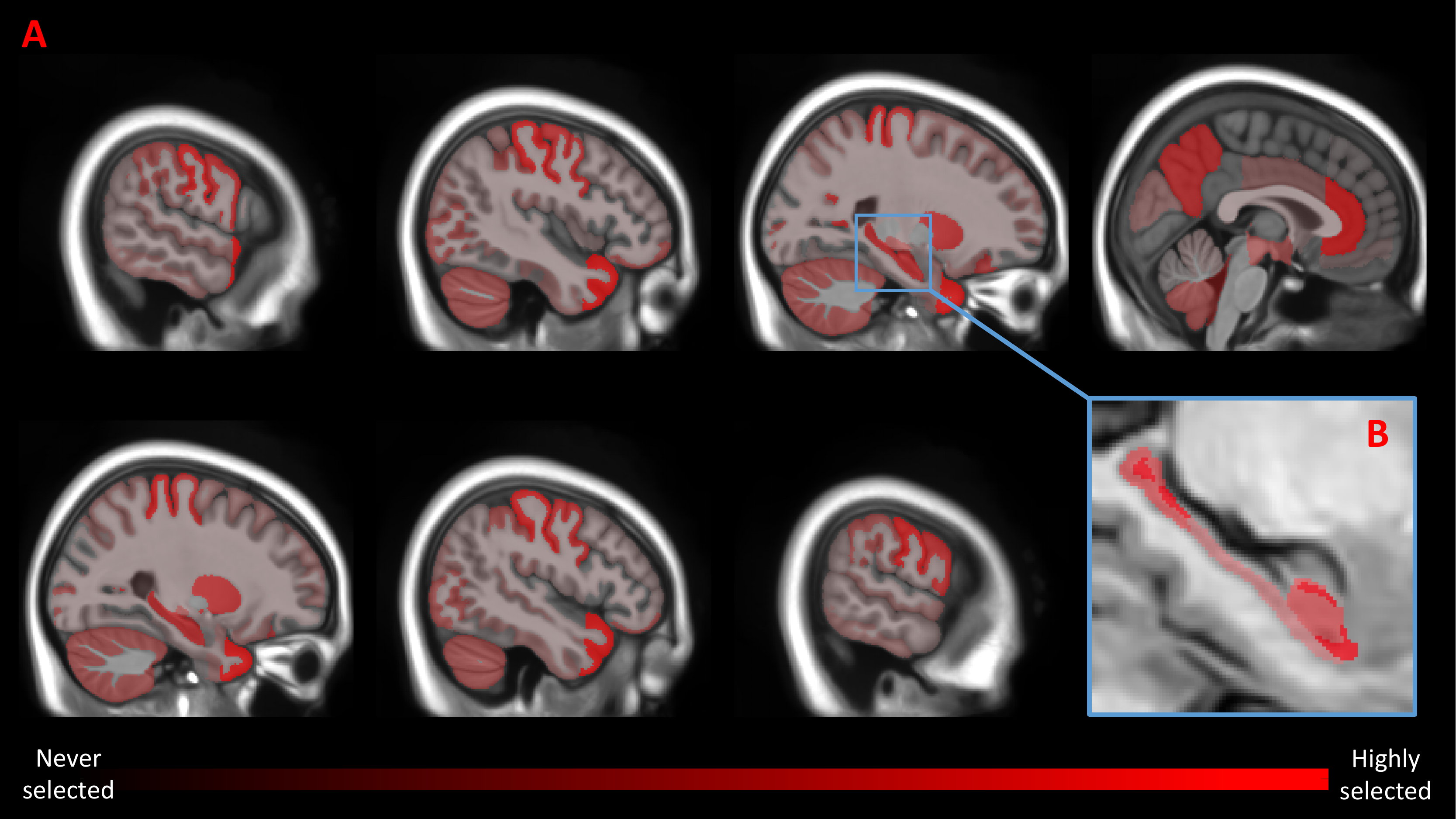}
\caption{Representation of the most selected structures. The brain structures and hippocampal subfields are selected separately with the elastic net method. Frequently selected structures are colored red. (A) the most frequently selected brain structures are the temporal lobe, the postcentral gyrus, the anterior cingulate gyrus, the hippocampus and the precuneus. (B) the most frequently selected hippocampal subfields are the CA1-SP, the CA1-SRLM, and the subiculum.} \label{fig:freqMap}
\end{figure*} 

First, we compared each element of our graph of structure grading within the hippocampal subfields (see Table~\ref{tab:results1}). As previously proposed in \citep{hett2018adaptive}, the PBG applied within the whole hippocampus is used as baseline for this experiment.

Thus, PBG based on the whole hippocampus structure obtains 76.8\% of AUC, 70.3\% of ACC and is more specific than sensitive. Although PBG values of all hippocampal subfields (see ``all'' in the table~\ref{tab:results1}) do not improve prediction performances, PBG values within selected vertices (\emph{i.e.}, subiculum, CA1-SP, and CA1-SRLM) obtain 77.1\% of AUC, 71.1\% of ACC (see ``selected'' in the table~\ref{tab:results1}), and improve the specificity in comparison to hippocampus grading. Thus, the use of hippocampal subfields selected with the elastic net method slightly increases the prediction performance of AD compared to the union of all subfields or the whole hippocampus. Furthermore, the edges selected by the elastic net do not  improve the prediction performance compared to other hippocampal features. Finally, the proposed method combining edges and vertices improves the AUC by 1.4 percent points and the accuracy 4.4 percent points compared to the global hippocampus grading. Our graph-based method also improves the AUC by 1.1 percent points and the accuracy by 3.6 percent points when compared to the use of the most discriminant hippocampal subfields. Moreover, in both cases, our proposed graph-based method has a higher sensitivity. 

The figure~\ref{fig:freqMap}-B illustrates the contribution (\emph{i.e.}, the number of selection by the elastic net) for each hippocampal subfield in the graph-based features vectors after the feature selection step. The experiments have shown that the most discriminant hippocampal subfields selected are the subiculum, and the two subfields representing the CA1. This is interesting because hippocampal subfields selected by EN method are in line with previous studies which have shown that the CA1 and subiculum as the subfields with the most significant atrophy in late stages of AD \citep{kerchner2012hippocampal,trujillo2014early}.

\begin{table*}
\centering
\caption{Classification of sMCI versus pMCI. Results obtained by inter-subject similarity features (\emph{i.e.}, vertices), intra-subject variability features (\emph{i.e.}, edges) and a combination of both. The patch-based grading applied on the hippocampus is used as baseline. The results demonstrate the genericity of our method that obtains best performances within the hippocampal subfields and the whole brain structures. Moreover, the experiments also show a slight superiority of the whole brain structures for AD prediction.}
\begin{tabular}{@{\hspace{0.7cm}} l @{\hspace{0.8cm}} c @{\hspace{0.8cm}} c @{\hspace{0.8cm}} c @{\hspace{0.8cm}} c @{\hspace{0.8cm}} c @{\hspace{0.7cm}}} \hline
Methods & AUC & ACC & BACC & SEN & SPE \\ \hline
Hippocampus PBG 	 & 76.8$\pm$0.2 	 & 70.3$\pm$0.0  	  & 70.6$\pm$0.0 		& 69.0$\pm$0.0 		&  72.2$\pm$0.0 	\\ \hline
Hipp. all vertices 	 & 73.9$\pm$0.2 	 & 67.1$\pm$0.0  	  & 67.9$\pm$0.0 		& 72.2$\pm$0.0 		& 63.5$\pm$0.0 		\\
Hipp. selected vertices  & 77.1$\pm$0.2 	 & 71.1$\pm$0.4  	  & 71.4$\pm$0.4 		& 69.5$\pm$0.6 		& \textbf{73.2$\pm$0.5} \\
Hipp. all edges		 & 66.7$\pm$0.2 	 & 61.1$\pm$0.4		  & 62.0$\pm$0.4		& 68.0$\pm$0.4		& 56.0$\pm$0.4 		\\
Hipp. selected edges 	 & 67.9$\pm$0.2 	 & 63.0$\pm$0.4  	  & 63.8$\pm$0.4  		& 68.9$\pm$0.4   	& 58.7$\pm$0.4  	\\
$GG_{subfields}$ 			 & 78.2$\pm$0.2 	 & 74.7$\pm$0.4  	  & 74.3$\pm$0.5 		& 77.1$\pm$0.5 		& 71.4$\pm$0.9		\\ \hline 
Brain all vertices 	 & 68.2$\pm$0.2 	 & 65.3$\pm$0.4  	  & 66.7$\pm$0.5  		& 68.6$\pm$0.5		& 62.2$\pm$0.5 		\\
Brain selected vertices  & 77.2$\pm$0.2 	 & 70.1$\pm$0.4  	  & 71.1$\pm$0.5  		& \textbf{77.8$\pm$0.5} & 64.4$\pm$0.5		\\
Brain all edges		 & 67.1$\pm$0.2		 & 65.7$\pm$0.2		  & 64.8$\pm$0.7		& 69.4$\pm$0.2		& 60.5$\pm$0.2 		\\
Brain selected edges     & 76.9$\pm$0.2 	 & 72.2$\pm$0.4  	  & 71.9$\pm$0.5  		& 73.8$\pm$0.5 		& 70.0$\pm$0.5	 	\\
$GG_{brain}$			 & \textbf{79.4$\pm$0.2} & \textbf{75.5$\pm$0.4}  & \textbf{75.1$\pm$0.5} 	& 77.6$\pm$0.5 		& 72.6$\pm$0.5 		\\ \hline
\end{tabular}
\label{tab:results1}
\end{table*}

\subsection{Graph of brain structures}

As done for hippocampal subfields, an evaluation of structure grading over the whole brain has also been conducted. We estimated the performance obtained by each feature separately (see Table~\ref{tab:results1}). The use of all vertices (\emph{i.e.}, the averages of PBG values computed within each brain structure) decreases the prediction performance compared to the use of only the hippocampus (65.3\% compared to 70.3\% of accuracy). A selection of the most discriminating vertices obtains similar results to those of the hippocampus only with an accuracy of 70.1\%. Contrary to the hippocampal subfields where vertices were most efficient than edges, the use of edge features performs similarly to the vertices.


Finally, as shown with the hippocampal subfields, the combination of both features, edges and vertices, that capture the inter-subjects' similarities and intra-subject variability enables an important increase of prediction performance. Our method applied with the brain structures obtains 75.5\% accuracy and 79.4\% AUC. Moreover, the experiments also show a similar sensitivity similar to that of using only selected vertices and a higher specificity than using only selected edges.

Figure~\ref{fig:freqMap}-A illustrates the most selected brain structures during the feature selection step. The experiments have shown that the the most frequently selected brain structures are the temporal lobe, the postcentral gyrus,  the anterior cingulate gyrus, the hippocampus and the precuneus. It is also interesting, as the results obtained from the hippocampal subfields, the most selected brain structures are in line with clinical studies that show a relationship between the atrophy of specific brain structures.


\begin{table*}[!ht]
\begin{center}
\caption{Comparisons of the different PBG approaches for the task of classifying sMCI versus pMCI. PBG computed over the hippocampus is provided as a baseline. The results show that the $MGG$ approach improves performance in terms of AUC, ACC, BACC, SEN and SPE. All results are expressed in terms of percentages.} \label{tab:results_multigraph}
 \begin{tabular}{@{\hspace{0.25cm}} l @{\hspace{0.3cm}} c @{\hspace{0.3cm}}  c @{\hspace{0.3cm}} c @{\hspace{0.3cm}} c @{\hspace{0.3cm}} c @{\hspace{0.25cm}}} \hline 
 Methods & AUC & ACC & BACC & SEN & SPE \\ \hline
 Hippocampus PBG & 76.8$\pm$0.2 & 70.3$\pm$0.0 & 70.6$\pm$0.0 & 69.0$\pm$0.0 &  72.2$\pm$0.0 \\ 
 Graph of hippocampal subfields ($GG_{subfields}$)  & 78.2$\pm$0.2 & 74.7$\pm$0.4 & 74.3 $\pm$0.4 & 77.1$\pm$0.4 & 71.4$\pm$0.4  \\
 Graph of brain structures ($GG_{brain}$) & 79.4$\pm$0.2 & 75.5$\pm$0.4 & 75.2 $\pm$0.4 & 77.6$\pm$0.4 & 72.6$\pm$0.4  \\
 Graph of hipp. sub. + brain & 79.6$\pm$0.2 & 74.5$\pm$0.4 & 73.9 $\pm$0.4 & 77.3$\pm$0.4 & 70.6$\pm$0.4  \\
 Multi-scale graph-based grading ($MGG$)*  & \textbf{80.6$\pm$0.2} & \textbf{76.0$\pm$0.4} & \textbf{75.7$\pm$0.4} & \textbf{77.8$\pm$0.4} & \textbf{73.6$\pm$0.4}  \\ \hline
 \end{tabular} 
 \end{center}\hfill\small{* Method illustrated in Figure~\ref{fig:multiple_graph}}
\end{table*}

\subsection{Multiscale graph-based grading}

A comparison of prediction performances obtained with our graph-based grading method applied in each anatomical scale independently and the combination of both is provided in Table~\ref{tab:results_multigraph}. In this experiment, two approaches have been investigated.

First, the results of this comparison confirm that for sMCI versus pMCI classification, whole-brain analysis enables better performance than analysis of the hippocampus subfields. Indeed, $GG_{brain}$ (whole brain) obtains 79.4\% of AUC and 75.5\% of accuracy while $GG_{subfields}$ (hippocampus subfields) obtains 78.2\% of AUC and 74.7\% of accuracy. 

Second, we compare the two approaches of combining both anatomical scales (\emph{i.e.}, simple concatenation or cascade of classifiers). The results suggest that the straightforward concatenation of the feature vectors from $GG_{brain}$ and $GG_{subfields}$ methods does not improve the performance compared to $GG_{brain}$ and $GG_{subfields}$. Indeed, the concatenation of the feature vectors obtains 79.6\% of AUC and 74.5\% of ACC, which is lower than the results obtained from the use of whole brain structures. However, the multi-scale graph-based approach ($MGG$) method (see Figure~\ref{fig:multiple_graph}) shows increased performance for each considered measure of classification. This last method obtains 80.6\% of AUC and 76\% of accuracy. This result indicates that the analysis of hippocampal subfields and whole brain structures are complementary. Therefore, in the rest of the experiments, we only consider the $MGG$ method.

\subsection{Complementarity with cognitive tests}

Table~\ref{tab:cognitivescore_multigraph} presents a comparison of the results obtained using features derived from cognitive tests, our imaging-based method and the combination of both. This comparison demonstrates that our imaging-based method obtains better results than using cognitive scores. Indeed, $MGG$ improves the sMCI versus pMCI classification by 1.8 percent of AUC and 1.5 percent of accuracy compared to using cognitive scores only. 

Moreover, the results of the experiment indicate the complementarity of imaging-based and cognitive assessments for AD prediction. Thus, the combination of cognitive scores and $MGG$ features obtains 85.5\% AUC and 80.6\% accuracy which improves AUC by 4.9\% and improves accuracy by 4.6\% when compared to the $MGG$ method.

\begin{table*}[!ht]
\centering
\caption{Comparison of our graph-based approach with cognitive test scores and combination of both for AD prediction. Although our MSGG obtains better results in terms of AUC, ACC, BACC, and SPE, the results of this comparison demonstrate the complementarity of our imaging-based method with cognitive scores. All results are expressed as percentage.} \label{tab:cognitivescore_multigraph}
 \begin{tabular}{@{\hspace{0.4cm}} l @{\hspace{0.4cm}} c @{\hspace{0.4cm}}  c @{\hspace{0.4cm}} c @{\hspace{0.4cm}} c @{\hspace{0.4cm}} c @{\hspace{0.4cm}}} \hline 
 Methods & AUC & ACC & BACC & SEN & SPE \\ \hline
 Cognitive score    & 78.8$\pm$0.2 & 74.5$\pm$0.4 & 72.4$\pm$0.4 & 84.9$\pm$0.4 & 60.0$\pm$0.4  \\
$MGG$  & 80.6$\pm$0.2 & 76.0$\pm$0.4 & 75.7$\pm$0.4 & 77.8$\pm$0.4 & \textbf{73.6$\pm$0.4}  \\
$MGG$ + Cognitive score & \textbf{85.5$\pm$0.2} & \textbf{80.6$\pm$0.4} & \textbf{79.2$\pm$0.4} & \textbf{87.3$\pm$0.4} & 71.1$\pm$0.4  \\ \hline
 \end{tabular}
\end{table*}

\subsection{Comparison with state-of-the-art methods}

Comparison with state-of-the-art methods is divided into two sub-comparisons. First, a comparison of MRI-based methods using similar ADNI datasets is provided in Table~\ref{tab:comparisons1_multigraph}. Second, a comparison of multi-modal methods was conducted. Besides cognitive assessments, the presented methods involved cerebral spinal fluid biomarkers (CSF), positron emission tomography (PET), and fluorodeoxyglucose PET (FDG-PET). The results of this comparison are provided in Table~\ref{tab:MultiSourceClass}.

\begin{table*}[!ht]
\begin{center}
\caption[Comparison with state-of-the-art methods]{Comparison with state-of-the-arts methods for Alzheimer's disease classification using similar ADNI1 dataset. In addition to sMCI versus pMCI, we provided results of CN versus AD classification. All results are expressed in percentage of accuracy (ACC) and balanced accuracy (BACC). Best balanced accuracy for each comparison is presented in bold font.}\label{tab:comparisons1_multigraph}
\begin{tabular}{@{\hspace{0.15cm}} l @{\hspace{0.3cm}} c @{\hspace{0.1cm}}c @{\hspace{0.1cm}}c @{\hspace{0.1cm}}c @{\hspace{0.3cm}} 
c @{\hspace{0.3cm}} c @{\hspace{0.3cm}} c @{\hspace{0.15cm}}} \hline
Methods 			 								&  \multicolumn{4}{c}{Subjects} 		& CN vs. AD 		& sMCI vs. pMCI \\ 
													& CN & sMCI & pMCI & AD	  & ACC (BACC) in \% 	& ACC (BACC) in \% 
													\\  \hline
Patch-based grading \citep{coupe2012scoring} 		& 231 & 238 & 167 & 198	& 88.0 (87.5) & 71.0 (71.0) \\
Sparse ensemble grading \citep{liu2012ensemble}		& 229 & $n.a$ & $n.a$ & 198	& 90.8 (90.5) & $n.a$ ($n.a$) \\ 
Voxel-based morphometry \citep{moradi2015machine} 	& 231 & 100 & 164 & 	200	& $n.a$ ($n.a$) & 74.7 (70.2)\\ 
Sparse-based grading \citep{tong2017novel} 		 	& 229 & 129 & 171 & 191 & $n.a$ ($n.a$) & 75.0 ($n.a$) \\
Multiple ensemble learning \citep{tong2014multiple} 	& 231 & 238 & 167 & 198	& 89.0 (89.5) & 70.4 (71.5) \\
Deep ensemble learning \citep{suk2017deep} 		 	& 226 & 226 & 167 & 186	 & 91.0 (91.3) & 74.8 (74.9)  \\ 
Hierarchical convolutional network \citep{lian2018hierarchical} 
													& 229 & 226 & 167 & 	199	& 90.3 (89.4) & 80.9 (69.0) \\
Deep neural network \citep{basaia2018automated}		& 352 & 510 & 253 & 295	& \textbf{99.2 (99.2)} & 75.1 (75.0) \\ 
Cortical graph neural network \cite{wee2019cortical}& 242 & $n.a$ & $n.a$ & 355 & 85.8 (85.5) & $n.a$ ($n.a$)\\
Proposed method										& 213 & 90 & 126 & 130 & 91.6 (91.4) & \textbf{76.0 (75.7)} \\ \hline
\end{tabular}
\end{center}
\end{table*}

The first comparison of $MGG$ results with state-of-the-art methods is provided in Table~\ref{tab:comparisons1_multigraph}. $MGG$ is compared with state-of-the-art methods using a similar ADNI1 dataset. Our graph-based method is compared with the original PBG method \citep{coupe2012scoring}, a graph-based grading method \citep{tong2014multiple}, an ensemble grading method \citep{liu2012ensemble}, a sparse-based grading method \citep{tong2017novel}, a VBM method \citep{moradi2015machine} and  advanced approaches based on deep ensemble learning technique \citep{suk2017deep,lian2018hierarchical,basaia2018automated,wee2019cortical}. 

The results of these comparisons demonstrate the competitive performance of our $MGG$ method for CN versus AD and sMCI versus pMCI classifications. Indeed, our method obtains state-of-the-art results with 91.6\% of accuracy for CN versus AD which are comparable to the most recent method based on deep-learning techniques. Furthermore, our method also obtains state-of-the-art performances for sMCI versus pMCI classification with 76.0\% of accuracy. These results are competitive with recent approaches based on deep-learning methods \citep{suk2017deep,lian2018hierarchical,basaia2018automated,wee2019cortical}. Moreover, our multi-scale graph-based method improves accuracy by 3.6 and 5 percentage points of the original PBG method for CN versus AD and sMCI versus pMCI classification, respectively \citep{coupe2012scoring}.

Moreover, as presented in Table~\ref{tab:MultiSourceClass}, our combination of $MGG$ and cognitive scores has been compared with state-of-the-art multimodal approaches. This comparison includes a method combining structural MRI and cognitive scores that obtains 80.7\% of accuracy \cite{tong2017novel}, a method combining MRI, PET scans and CSF that obtains 83.3\% of accuracy \cite{suk2015latent}, a voxel-wise approach that combines MRI, FDG-PET and cognitive scores that obtains 80.9\% of ACC \cite{samper2019reproducible}, and a recent multimodal deep-learning approach combining MRI, CSF and cognitive scores that obtains 76 \% of accuracy \cite{lee2019predicting}. This demonstrates the competitive performance of our graph-based approach that obtains state-of-the-art results with only the use of MRI-based and cognitive score features.

\begin{table*}[!ht]
\begin{center}
\caption{Comparison of the different combination of different imaging biomarkers CSF, and cognitive scores used in clinical routines for the prediction of MCI conversion (\emph{i.e.}, sMCI versus pMCI comparison). All results are expressed as percentage. Best AUC is expressed in bold font.} \label{tab:MultiSourceClass}
 \begin{tabular}{@{\hspace{0.4cm}} l @{\hspace{0.4cm}} c @{\hspace{0.4cm}}  c @{\hspace{0.4cm}} c @{\hspace{0.4cm}}} \hline 
 Methods 	& Source 		 & AUC 		& ACC  \\ \hline
 Latent feature representation \cite{suk2015latent} & MRI + PET + CSF & n.a & 83.3\\
 Combined sparse-based grading \cite{tong2017novel} & MRI + Cognitive scores$^1$ & 87.0 & 80.7 \\
 Voxel-wise approach \cite{samper2019reproducible} & MRI + FDG-PET + Cognitive score$^2$ & \textbf{88.5} & \textbf{80.9} \\
Multimodal deep learning approach \cite{lee2019predicting} & MRI + CSF + Cognitive scores$^3$ &  n.a & 76.0 \\
Proposed & MRI + Cognitive scores$^4$  & 85.5 	& 80.6  \\ \hline
 \end{tabular}
 \end{center}
 $^1$ Age, MMSE, CDR-sb, RAVLT, ADAS\\ 
 $^2$ Gender, MMSE, Education level, CDR-sb, RAVLT, ADAS \\
 $^3$ ADNI-EF, ADNI-MEM\\ 
 $^4$ MMSE, CDR-sb, RAVLT, FAQ, ADAS11, ADAS13\\ 
\end{table*}

\section{Discussion}

The first contribution of this paper is the development of a new graph-based grading approach that combines inter-subject similarities and intra-subject variability efficiently. We validated this new method with two different anatomical scales: the hippocampal subfields and the whole brain structures. The second contribution is the development of an anatomical scale fusion based on a cascade of classifiers approach. We applied this multi-scale graph-based grading framework to the hippocampal subfields and a parcellation of the entire brain structures. To validate our new multi-scale graph-based grading framework, we compared each component of our graph at each anatomical scale. Then, we compared the results obtained in our experiments with the results of state-of-the-art methods proposed in the literature. Finally, we compared the results obtained with our imaging-based biomarker with a bank of cognitive scores that are used in clinical routines.

\subsection{Graph of hippocampal subfields}

Postmortem and \emph{in-vivo} studies have suggested that the first regions of the brain which are changed in typical disease progression are the entorhinal cortex (EC) and the hippocampus \citep{jack1992mr,braak1995staging,bobinski1999mri}. Moreover, neuroimaging studies have shown that the hippocampus is the brain structure with the most significant alterations at the early stage of AD \citep{frisoni2010clinical,schwarz2016large}. However, recent methods applied to the hippocampus have shown limited performances for AD prediction \citep{hett2017adaptive,tong2017novel}. This limitation could come from global analysis of the hippocampus. Indeed, the hippocampus is subdivided into several subfields. The terminology differs across segmentation protocols \citep{yushkevich2015quantitative} but the most recognized definition \citep{lorente1934studies} mainly divides the hippocampus into the subiculum, the cornu ammonis (CA1/2/3/4), and the dentrate gyrus (DG). Furthermore, studies showed that hippocampal subfields are not equally impacted by AD \citep{braak1997alzheimer,braak2006staging,apostolova2006conversion,la2013hippocampal,kerchner2010hippocampal,kerchner2012hippocampal}. Indeed, postmortem, animal-based and recent \emph{in-vivo} imaging studies showed that the CA1 and the subiculum are the subfields impacted by the most discriminant atrophy in the last stage of AD \citep{apostolova2006conversion,la2013hippocampal,kerchner2012hippocampal,li2013discriminative,trujillo2014early}. This indicates that the analysis of the hippocampus with a global measure could limit prediction performance. Indeed, better modeling of the structural alterations within the hippocampal subfields could improve prediction performance.

Consequently, we proposed to better model hippocampus alterations with the application of our novel graph-based framework within the hippocampal subfields. First, we studied the efficiency of a straightforward approach that computes the average of grading values in each hippocampus subfield separately instead of the whole hippocampus structures as is usually done. This results in poorer performance compared to the average of grading values within the whole hippocampus. However, the grading values within the most discriminant hippocampal subfields (\emph{i.e.}, subiculum and the two definitions of CA1) obtain similar performances to the average of grading values within the whole hippocampus. This might be due to the fact that the subiculum and CA1 represent the major part of the hippocampus. 

The related hippocampal subfield features selected by the elastic net are consistent with previous \emph{in-vivo} imaging studies, which are based on 3T MRI and ultra-high field MRI at 7T. These studies analyzed the atrophy of each hippocampal subfield at an advanced stage of AD. These studies showed that CA1 is the subfield with the most severe atrophy \citep{apostolova2006conversion, mueller2007measurement, la2013hippocampal, carlesimo2015atrophy}, and also indicate that CA1SR-L-M is the subfield with the greatest atrophy at advanced stages of AD \cite{kerchner2010hippocampal,kerchner2012hippocampal}. It is interesting to note that the results of our experiments are also in accordance with previous postmortem, animal-based, and \emph{in-vivo} studies combining volume and diffusivity MRI. These last studies demonstrated that the subiculum is the earliest hippocampal region affected by AD \citep{trujillo2014early, li2013discriminative}.

Finally, the great improvement obtained with the combination of inter-subject similarities and intra-subject variability shows that this information is complementary. It also confirms this combination enables to obtain results similar to methods based on whole brain analysis with only the use of the hippocampus.

\subsection{Graph of brain structures}

Next, we investigated our method at the whole brain scale. The comparison of hippocampus PBG and the most discriminant vertices indicate that the straightforward combination of other discriminant brain structures does not increase the prediction performance compared to using only the hippocampus. Moreover, when the edges and the vertices are combined, our experiments show that the edges are the most discriminant selected elements. 

Our experiments indicate that the most selected brain structures are the postcentral gyrus, the anterior cingulate gyrus, the hippocampus, and the precuneus (see Figure~\ref{fig:freqMap}). These results are in line with the literature. First, it is interesting to note that the automatic selection of most discriminant feature shows the importance of the temporal lobe and the hippocampus. Studies have shown a significant loss of gray matter within the temporal lobe \citep{killiany1993temporal,busatto2003voxel}, while the hippocampus has long been known as the structure with the earliest alterations \cite{hyman1984alzheimer,west1994differences,braak1995staging,ledig2018structural}. Second, VBM and perfusion studies have shown that the precuneus suffers from a noticeable atrophy and a bilateral decrease of regional cerebral blood flow compared to control subjects \cite{kogure2000longitudinal,karas2007precuneus}. Studies have shown a significant reduction in volume of the anterior cingulate gyrus compared to control \cite{frisoni2002detection,jones2006differential}. Moreover, a study showed that the volume of the anterior cingulate gyrus is correlated with apathy which is symptomatic of AD \cite{apostolova2007structural}. However, the importance of the postcentral gyrus was unexpected since it has been shown that this structure seems unaffected by AD process \cite{halliday2003identifying}. These elements seem to indicate that the structural pattern of AD is composed of both highly impacted and healthy brain structures.

Finally, the good performance of the graph-based grading method demonstrates that the combination of both features enables a better discrimination of subjects who convert to dementia in the years following their first visits. Moreover, the good results within hippocampal subfield and brain structure parcellation indicate the generic nature of our new method, which can be applied at different anatomical representation.

\subsection{Multi-scale graph-based grading}

Afterwards, we compared the results of our multi-scale ($MGG$) approach with the previously described $GG_{brain}$ and $GG_{subfields}$. First, the conducted experiments show that our graph of structure grading applied within hippocampal subfields improves prediction of conversion to Alzheimer's disease compared to the PBG applied within the hippocampus. 

The results obtained by the straightforward extension of the graph of structure grading to combine whole brain structure and hippocampal subfields did not demonstrate an improvement in AD conversion prediction compared to the single use of $GG_{brain}$ and $GG_{subfields}$. The main limitation might come from the fact that the straightforward combination of different anatomical representations suffer from a substantial augmentation of feature dimensionality. To address these limitations, we proposed the $MGG$ method that is based on a cascade of classifiers. This method alleviates the dimensionality issue by estimating an intermediate conversion probability for each anatomical scale considered. This results in an increase in AD prediction performances compared to $GG_{brain}$ and $GG_{subfields}$ methods.

\subsection{Comparison with state-of-the-art methods}

In this last decade, many improvements in computer-aided diagnosis methods have been proposed to better capture structural alterations using anatomical MRI (see \cite{rathore2017review} for a review). Two main approaches have been proposed: methods based on inter-subject similarity \cite{coupe2012scoring,moradi2015machine,tong2017novel} and methods based on intra-subject variability \cite{tong2014multiple,suk2014hierarchical}. Consequently, the first contribution of our work was to combine  inter-subject similarity -- using the PBG framework -- and the intra-subject variability -- with the integration of PBG into a graph-based model. A strength of our method reside in its generic nature. Indeed, our graph-based grading can obtain competitive results with different anatomical representations.

Another difference with state-of-the-art methods comes from the proposition of a multi-anatomical scale analysis of AD alterations. In contrast to previous methods which analyzed changes at a unique anatomical scale (\emph{i.e.}, cortical cortex, whole brain structures, hippocampus, or hippocampal subfields,...), we proposed combining whole brain structures parcellation with a representation of hippocampal subfields. This combination has resulted in performances competitive with state-of-the-art methods.

Finally, the comparison with state-of-the-art approaches using similar ADNI1 subset has shown that our multi-scale graph-based grading method obtains competitive results for both AD detection and prediction. The high performances of the methods proposed in \cite{basaia2018automated} and \cite{lian2018hierarchical} have to be moderated. First, it is unclear how \citep{basaia2018automated} obtains high accuracy for AD detection since misdiagnosis of AD in ADNI dataset is around $4\%$. Second, the excellent result of \cite{lian2018hierarchical} for AD prediction mainly takes advantage of unbalanced results, after the accuracy is corrected (see Table~\ref{tab:comparisons1_multigraph}), this approach obtains lower prediction performance.

\subsection{Complementarity with cognitive tests}

Finally, an analysis of the complementarity of our imaging-based method with scores resulting from cognitive assessments has been carried out. These experiments enabled the comparison of the performance of cognitive scores and our imaging biomarker for AD prediction. 

First, the conducted experiments demonstrate that our graph-based grading approach using T1 weighted MRI obtains substantially better results for the prediction of AD than the single use of cognitive scores. Moreover, as shown in many works listed in Table~\ref{tab:MultiSourceClass} \cite{suk2015latent,tong2017novel,samper2019reproducible,lee2019predicting}, MRI-based biomarkers and cognitive assessments are complementary. Indeed, their combination improves classification performances. Thus, the combination of our graph-based grading technique and cognitive assessments demonstrates a great improvement in performance compared to the use of each method separately. This improvement is comparable to studies based on multimodality frameworks, which use more expensive biomarkers (\emph{e.g.}, PET, FDG-PET,...), are less reproducible (often due to sample storage issues), and are highly invasive (\emph{e.g.}, CSF).

\section{Conclusion}

Improved modeling of AD alterations is a great challenge that could lead to earlier predictions of conversion. Therefore, in this work, we developed a new method to better model AD signature. Our proposed method models the pattern of AD alterations by combining inter-subject similarity and intra-subject variability. The conducted experiments have shown the generic nature of our new framework. Consequently, we proposed a multi-anatomical scale graph-based grading method to combine the alterations at different anatomical scales. In addition, we conducted the first joint analysis of the hippocampus subfields and brain structure changes in the same framework. The results show state-of-the-art-performance, confirming the complementarity of hippocampal subfields and whole brain analysis, and the complementarity of inter-subject similarity and intra-subject variability.

\section{Acknowledgement}
This work benefited from the support of the project DeepvolBrain of the French National Research Agency (ANR-18-CE45-0013). This study was achieved within the context of the Laboratory of Excellence TRAIL ANR-10-LABX-57 for the BigDataBrain project. Moreover, we thank the Investments for the future Program IdEx Bordeaux (ANR-10- IDEX- 03- 02, HL-MRI Project), Cluster of excellence CPU and the CNRS.

Data collection and sharing for this project was funded by the Alzheimer's Disease Neuroimaging Initiative (ADNI) (National Institutes of Health Grant U01 AG024904) and by the National Institute on Aging, the National Institute of Biomedical Imaging and  Bioengineering, and through generous contributions from the following: AbbVie, Alzheimer's Biogen; Bristol-Myes Squibb Company; CereSpir, Inc.; Cogstate; Eisai Inc.; Elan Pharmaceuticals, Inc.; Eli Lilly and Company; EuroImmun; F. Hoffman-La Roche Ltd and its affiliated company Genentech, Inc.; Fujirebio; GE Healthcare; IXICO Ltd.; Janssen Pharmaceutical Research \& Development LLC.; NeuroRx Research; Neurotrack Technologies; Novartis Pharmaceuticals Corporation; Pfizer Inc.; Piramal Imaging; Servier; Takeda Pharmaceutical providing funds to support ADNI clinical sites in Canada. Private sector contributions are facilitated by the Foundation for the National Institutes of Health (\url{www.fnih.org}). The grantee organization is the Northern California Institute of Research and Education, and the study is coordinated by the Alzheimer's Therapeutic Research Institute at the University of Southern California. ADNI data are disseminated by the Laboratory for Neuro Imaging at the University of Southern California.

\section{Conflict of interest}
The authors declare that the research was conducted in the absence of any commercial or financial relationships that could result as a potential conflict of interest.

\bibliographystyle{plain} 
\bibliography{ni18.bib}

\end{document}